%% file: main.tex
\documentclass[myepj-spec,pdftex]{mySvjour}
\usepackage{graphicx}
\usepackage[pdfpagemode=UseNone]{hyperref}
\usepackage{color}
\usepackage{xspace}
\usepackage[square,sort&compress,numbers]{natbib}   
\usepackage{verbatim}
\usepackage{amsmath,amssymb,amsfonts} 
\usepackage{tabularx}
\usepackage{multicol}
\usepackage{footnote}
\usepackage{listings}
\usepackage[gen]{eurosym}
\usepackage[switch, modulo]{lineno}

\usepackage{lmodern,bm}                
\usepackage[T1]{sansmath} 
\SetMathAlphabet{\mathsfbf}{sans}{\sansmathencoding}{\sfdefault}{bx}{sl}
\usepackage{etoolbox}
\AtBeginEnvironment{sansmath}{}{}{}

\usepackage{lipsum}

\newcommand{\June}{\textsc{June}\xspace}
\newcommand{\JuneGermany}{\textsc{June-Germany}\xspace}

\definecolor{darkblue1}{rgb}{0,0,.2}
\definecolor{darkblue}{rgb}{0,0,.2}
\definecolor{darkred}{rgb}{0.5,0,0}
\pagecolor{white} 
\color{black}     
\hypersetup{breaklinks=true, 
	colorlinks=true, 
	linkcolor=darkblue1, 
	menucolor=darkblue1, 
	urlcolor=darkblue1,
	citecolor=darkblue1,
	pdftitle={},
	pdfauthor={},
	pdfsubject={},
	pdfkeywords={},
	pdfproducer={}
}
\usepackage{siunitx}

\bibstyle{plain}

\begin{document}
	
	\twocolumn[{%
		\begin{@twocolumnfalse}
			
			\begin{flushright}
				\normalsize
			\end{flushright}
			
			\vspace{-2cm}
			
			\title{\Large\boldmath JUNE-Germany: An Agent-Based Epidemiology Simulation including Multiple Virus Strains, Vaccinations and Testing Campaigns}
			\input{author_list}
			
			\abstract{
The \June software package is an open-source framework for the detailed simulation of epidemics based on social interactions in a virtual population reflecting age, gender, ethnicity, and socio-economic indicators in England. In this paper, we present a new version of the framework specifically adapted for Germany, which allows the simulation of the entire German population using publicly available information on households, schools, universities, workplaces, and mobility data for Germany. Moreover, \JuneGermany incorporates testing and vaccination strategies within the population as well as the simultaneous handling of several different virus strains. First validation tests of the framework have been performed for the state of Rhineland Palatinate based on data collected between October 2020 and December 2020 and then extrapolated to March 2021, i.e. the end of the second wave.}	
	\maketitle
	\end{@twocolumnfalse}
}]

\tableofcontents

\section{Introduction}	

Modelling the spread of transmissible pathogenic diseases within a population has become relevant not only since the Covid-19 pandemic but also for pandemic preparedness for other potential future outbreaks, such as highly pathogenic influenza. It is therefore necessary to investigate what measures could be used to effectively contain an epidemic. Various modelling approaches have been developed in recent years and decades. These range from analytical models based on differential equations to compartmental models and purely data-driven parameterizations. An overview of the different approaches can be found in the following references \cite{articleOverview1, articleOverview2, inbookCompart}.\par
A particularly promising approach is agent-based models \cite{article4, Rockett2020, articleSocial}, where each individual in a population is simulated by software agents. The characteristics of the population relevant to the spread of infectious diseases, such as age, living and working conditions, or social context, are introduced into the simulation by external data, and the agents are constructed according to the actual statistical distributions. Depending on the infectious disease being simulated, transmission dynamics are also implemented. Free or undetermined parameters in the model can then be determined in a second step based on the observed data from the real pandemic, such as the number of hospital admissions. The major advantage of agent-based models (ABMs) is the ability to study different policy mitigation strategies when implemented in different geographical and social contexts. \par
The \June framework \cite{Bullock:2021} is an agent-based model for simulating epidemics in a population and was developed during the first and second waves of the Covid-19 pandemic. It incorporates highly granular geographic and sociological resolution for England, based on 2011 census data \cite{Zensus2011}. Notably, \June was the first model to predict the spread of COVID-19 with high geographical and sociological accuracy.\par
While the updated version of \June (v1.2) \cite{June12} allows for the simulation of vaccination and multiple viruses, this was not the case for the initial version (v1.0) on which \JuneGermany is based. We, therefore, developed an independent approach to handle testing strategies and vaccines for a broad population, as well as the simultaneous simulation of multiple virus variants.\par
In chapter \ref{sec:June} we first present the basic features of the original \June framework (v1.0) before discussing the functional extensions in chapter \ref{sec:extension}. The input data for Germany are summarised in chapter \ref{sec:germany}. The estimation of the model parameters is discussed in section \ref{sec:parameters}. The validation of the program was performed using data from the second wave of covid-19 between October 2020 and March 2021 ( Chapter \ref{sec:validation}) for the state of Rhineland-Palatinate. The paper concludes with a summary in chapter \ref{sec:conclusion}.

\section{Summary of the \June Framework v1.0 \label{sec:June}}

The \June framework is presented and discussed in detail in \cite{Bullock:2021}. As a result, we will only give a brief summary of the main features of the original framework. The features of the \June (v1.2) framework are discussed in Ref.~\cite{June12}. The framework is implemented in Python and is structured in four interconnected layers: The population layer contains the necessary information about individual agents and their static social environments, such as households and workplaces. The interaction layer describes the daily routines of the agents, e.g. their commute to work or school and their leisure activities. The disease layers model the characteristics of disease transmission and the effects on the agents within the simulation. Government policies to mitigate the effects of the pandemic are included in the policy layer of the framework.\par
The demographics of the virtual population are divided into hierarchical geographical layers, e.g. states/counties, districts/regions and/or boroughs/parishes. The age and gender distribution of each layer are known. Each individual agent in the simulation is a member of a household (e.g. single, couple, family) and lives with a specific known number of adults aged over 65, other adults, dependent adults and/or children. Students are associated with nearby schools and universities according to their age and location. Working members of a household are assigned to three employment categories, i.e. work in companies with employees, work outside fixed company structures and work in hospitals and schools. The distribution of working persons across workplaces depends on their location, social environment and sector of work. Hospitals are treated separately because of their particular importance during a pandemic. Social contact networks are constructed by linking each household to a list of other households that have a high probability weight of being nearby.\par
Each day an agent is simulated in discrete time steps of variable length, distinguishing between weekdays and weekends. For example, a typical weekday consists of 8 hours at work, 10 hours at home and the remaining time for other activities, e.g. meeting other agents in a defined context such as shopping, visiting restaurants, visiting friends or relatives at home, etc.. This varies for each agent according to age and sex.\par
It has been assumed that almost no disease transmission occurs during individual agents' car journeys, so the focus of the simulation is on public transport. Major transit cities are represented as nodes, defining a simplified model of a transport network. Commuting within cities and regions is modelled as a self-connected loop. Agents are placed in carriages of 30-50 agents when commuting within a city or region. Transmissions can then occur between agents within these carriages.\par
The frequency and intensity of further personal contact in different social settings is modelled by social interaction matrices, which are age-dependent. For example, pupils in schools have a high probability of meeting other pupils of the same age. The social interaction matrix actively depends on the risk reduction policy, e.g. school closures, applied at a given time step.\par
The \June framework uses probabilistic modelling of infection transmission from an infected and transmitting host to a susceptible host. The probability of being infected is assumed to be a statistical Poisson process. The probability of actual infection depends on several factors: The number of infectious agents present at a given location, the probability of transmission of the host at a given time, the susceptibility of the potential host, the exposure time interval during which a given group is in the same location, the number of possible contacts, the proportion of physical contacts, and the overall intensity of group contact at a given location. Several of these parameters can be impacted by specific mitigation, testing, and vaccination strategies. Once infected, hosts in the simulation will experience different health outcomes, ranging from asymptomatic individuals to those requiring hospital treatment and those requiring admission to intensive care.\par
Government policies can be implemented at a local level, taking into account geographical regions, nature, and the location of social interactions or workplaces. It can therefore be modelled on parts of the population who are essential workers and continue to go to work while the rest stay at home. In addition, the \June framework allows modelling the compliance of the general population with government policies. This again depends on various social and demographic parameters. 

\section{Features of the \JuneGermany Framework \label{sec:extension}}

The original \June framework was developed primarily during the first wave and also at the beginning of the second wave of the Covid-19 pandemic in England when only one type of virus was present. As a result, only a limited range of testing strategies had been implemented and no vaccinations were available for the general population. These three missing aspects of \June v1.0 were implemented as part of this work and are described in the following section. An updated version of \June (v1.2) for England has since become available, which allows modelling of similar features. 

\subsection{Test Strategies}

During the Covid-19 pandemic, several countries introduced testing policies and strategies. In Germany, for example, people living in high-incidence areas are required to show a negative test result before entering restaurants or leisure facilities. This form of rapid testing for asymptomatic agents is often carried out in city centres, near leisure facilities and also in schools. People who test positive during this process must be quarantined according to the policies of the federal or regional government at the time. Within the framework of \JuneGermany, testing strategies have now been implemented that allow these tests to change the infection status of agents at selected locations, e.g. schools, workplaces or shops. By specifying the days and times of the week when the tests are carried out, this simulation allows changes in the frequency of testing. False positives are not considered in the simulation. False negative cases are modelled by the sensitivity of the test depending on the symptoms of the infected person, indicating the likelihood of a negative result even though the agent is known to be infected. Agents who test positive will stay at home for the quarantine period specified in the policy, but this behaviour is subject to the social compliance factor of the model.

\subsection{Vaccinations}

The \JuneGermany framework allows modelling the vaccination of a given population with a vaccine (BioNTech). The population is divided into priority groups according to age and comorbidity. It is also possible to select specific workplaces to be prioritised. The vaccine distribution is modelled in terms of doses per day, which are then distributed among the priority groups. The increase in vaccine availability is modelled linearly per day, which is part of the vaccination strategy. The time between the first and second vaccination is 42 days by default and does not vary during the simulation. A person who has been infected within the last six months receives only one dose. The vaccine is parameterised by the reduction factors for the risk of infection, hospitalisation or death, and the reduced risk of onward transmission to other agents. We distinguish between the first and second doses to modify these factors. These factors also depend on the type of virus variant with which the agent may be infected. The vaccines are time-dependent from the time of vaccination, which means that they are only effective 14 days after either vaccination.\par

\subsection{Multiple Viruses}

In principle, the original \June (v1.0) framework allows the simulation of different virus types, but these would have to be run sequentially. As a result, the interaction between viruses would not be possible. However, the new variants of the SARS-CoV-2 virus that emerged during the COVID-19 pandemic need to be considered simultaneously in the simulation, as immunity of recovered agents to one virus type also provides some protection against infection with another virus type. The different viruses typically differ in their transmissibility, symptomatology and lethality. To simulate this in \JuneGermany, a new infection class was defined that takes these parameters into account and can be defined for any number of virus variants simultaneously in the simulation. Each infected agent is assigned an additional attribute that defines the variant of the virus. Then, when calculating the probabilities of infection within the interaction class for each group, the probability of each variant is considered separately.\par
The immunity of an agent after surviving infection by other virus variants is described by a vector. The effects of vaccination and recovery can also be adjusted to treat susceptibility to other virus variants differently. It is now possible to study different infectious diseases and their interactions simultaneously.\par 
In principle, it would also be possible to simulate completely different diseases, such as Covid-19 and influenza, at the same time. 


\section{Population and Properties of Germany \label{sec:germany}}

\subsection{Geography and demography}

The geographical model is based on the German administrative regions (from VG250) \cite{VG250}. The geographical model consists of three layers, the coarsest of which are the 16 German states. The next layer consists of 413 districts, which are further subdivided into a total of 11,564 municipalities. As the most detailed demographic dataset available is the 2011 Census (Zensus)~\cite{Zensus2011}, it is chosen as the geographical basis for this simulation. This dataset also includes a description of household composition.\par
\begin{figure*}[htb]
\begin{center}
\begin{minipage}[t]{0.49\textwidth}
\includegraphics[height=0.75\textwidth]{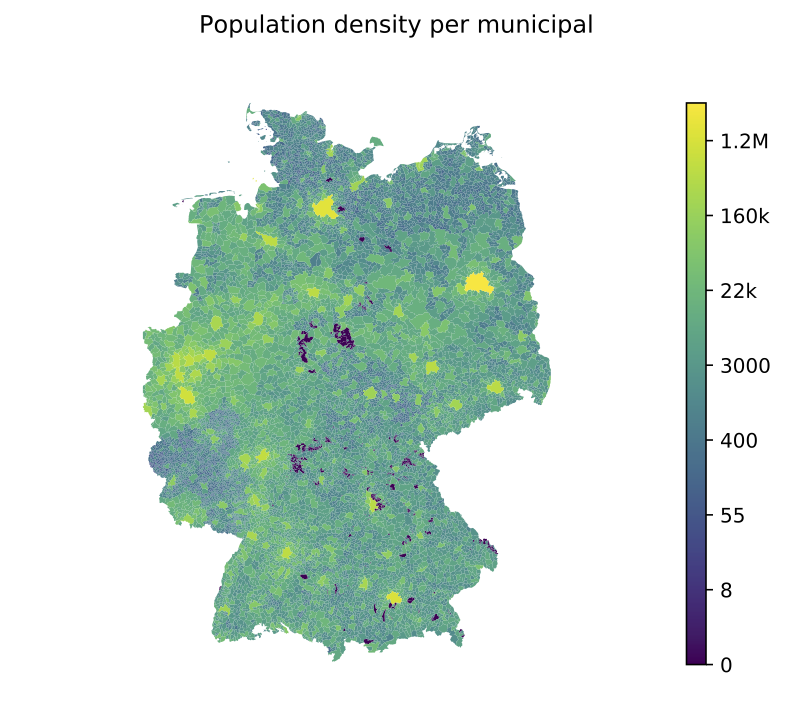}
\caption{\label{fig:population-density} Representation of the population density within Germany used in \JuneGermany.}
\end{minipage}
\hspace{0.1cm}
\begin{minipage}[t]{0.49\textwidth}
\includegraphics[height=0.75\textwidth]{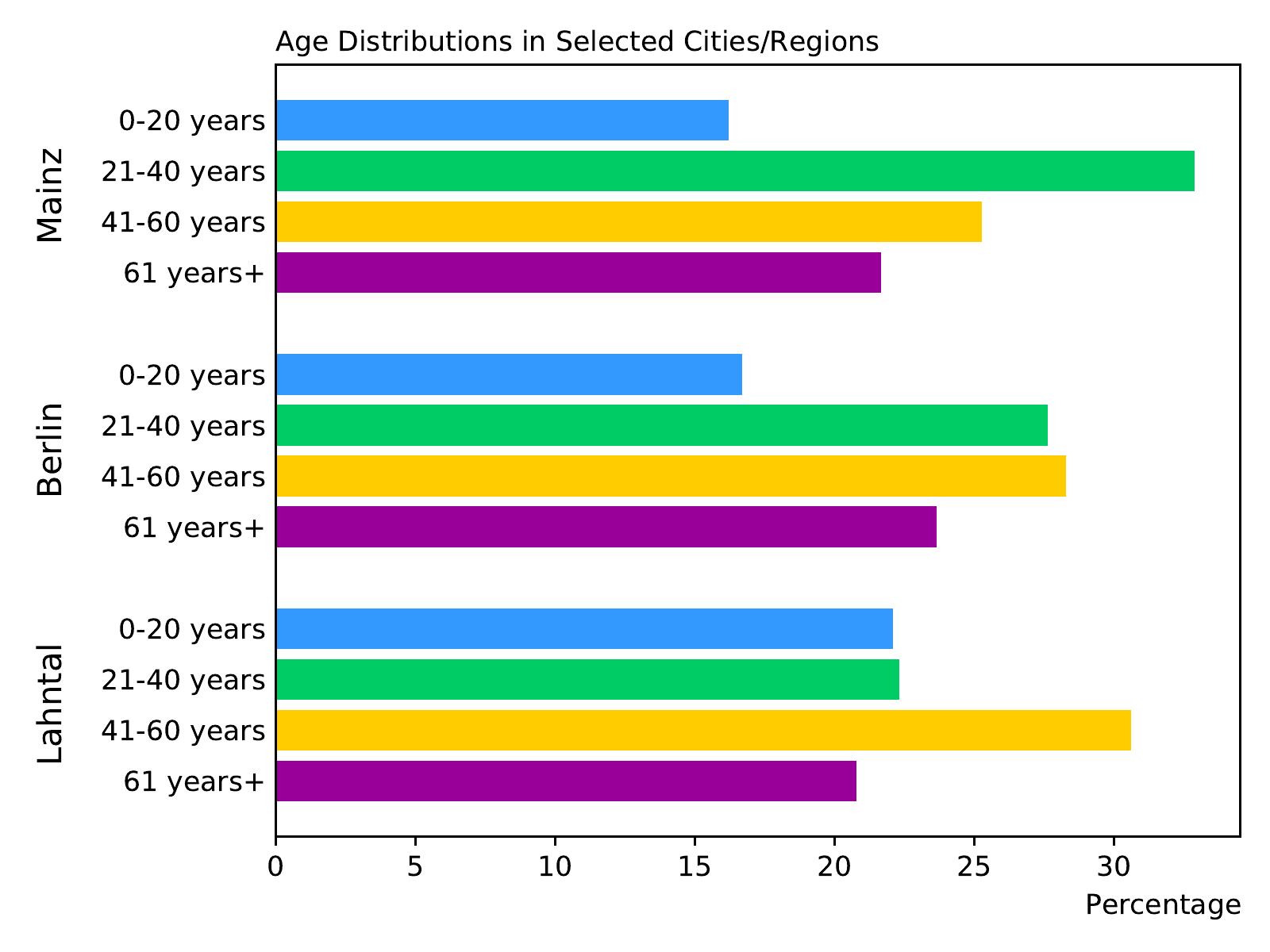}
\caption{\label{fig:population-selected} Overview of the summarized age distribution if three selected cities of different sizes.}
\end{minipage}
\end{center}
\end{figure*}

The population in each municipality is generated based on the age and sex distribution in age steps of one year between 0 and 100 years. The resulting population density at the municipality level is shown in \autoref{fig:population-density}. The most populated municipality is Berlin with 3.3 million residents, while the least populated municipality is Dierfeld (in the district of Bernkastel-Wittlich) with only 9 people. On average, each municipality contains 7073 persons, with a large spread ($\sigma = 44000$) due to some large cities are not divided into smaller municipalities. To illustrate the different age structures in cities, \autoref{fig:population-selected} shows rough age distributions for a medium-sized city (Mainz) with 200,000 inhabitants, the largest city (Berlin), and a small district (Lahntal) with less than 10,000 inhabitants. The visible differences between these three cities illustrate the importance of a fine-grained resolution of population statistics, as people of different age groups not only react differently to an infection but even more importantly behave differently due to different social interaction patterns.\par
The German population is generated according to official demographic data. Considerable effort has gone into collecting data to fill in the attributes of each person. We distribute the co-morbidity of the German population with the data for the English population. We are aware of the differences between these populations but argue that the overall effect on accuracy will be negligible. The most important risk factor for a severe Covid-19 case is age, which we model correctly. So the differences in comorbidity are negligible.

\subsection{Households}

Household compositions are extracted from the 2011 Census dataset \cite{Zensus2011} and translated to match the required input of the existing household distributor. Households are described by a limited number of categories depending on the number of adults and children living in each household. For single and couple households, a further distinction is made between middle-aged (21-60 years) and elderly (>60 years). For middle-aged households, a distinction is made between households with and without children or young adults, and the number of children/young adults also varies. There is a category for households with only students. As the 2011 census data set does not include the definition of student households, the number of students is estimated on the basis of the number of students per country and the number of persons aged 18-29 in each municipality.\par
\begin{figure*}[htb]
\begin{center}
\begin{minipage}[t]{0.49\textwidth}
\includegraphics[height=0.75\textwidth]{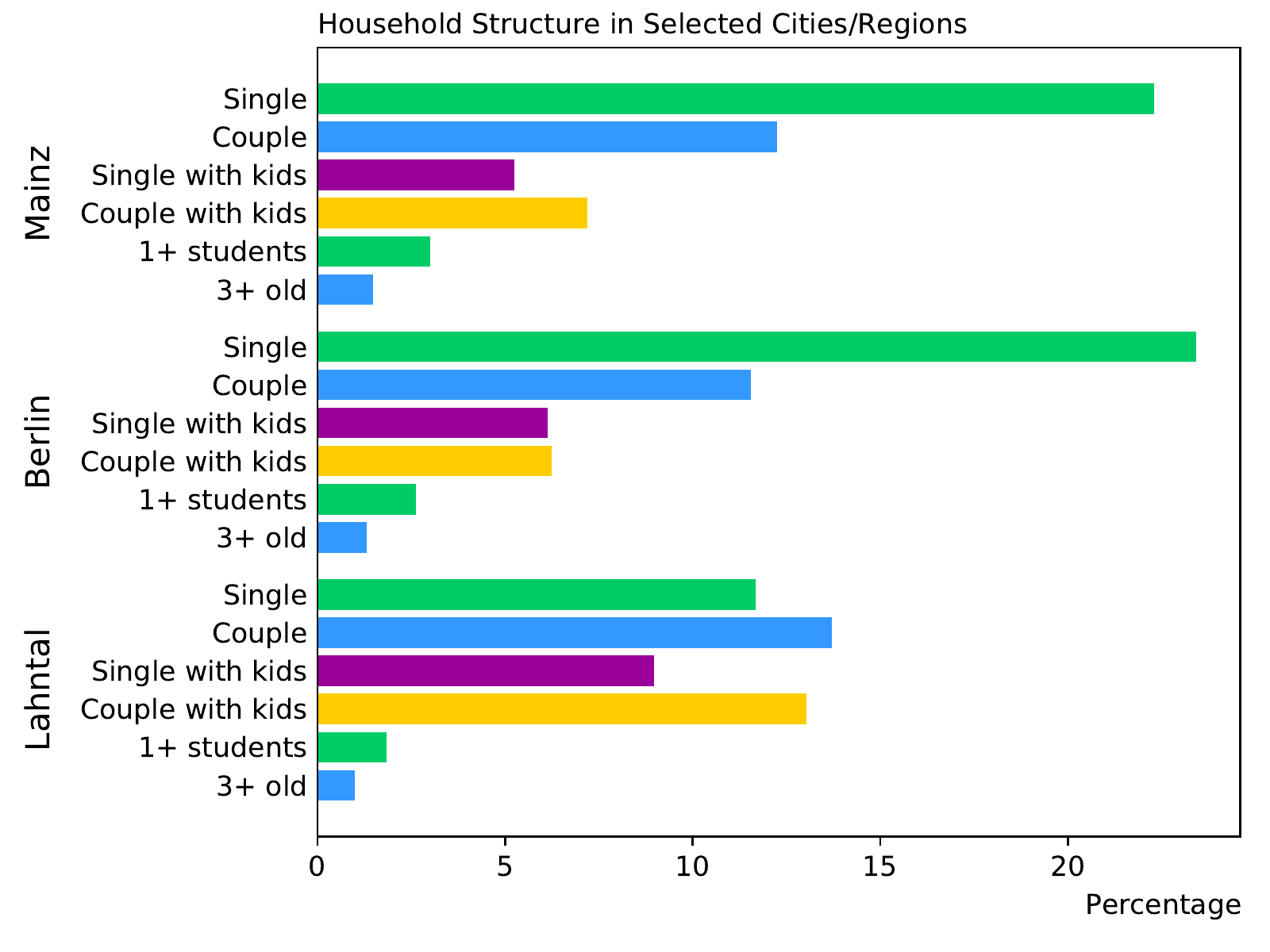}
\caption{\label{fig:households} Overview of the household distribution of three selected cities of different sizes. For improved readability, similar categories are merged.}
\end{minipage}
\hspace{0.1cm}
\begin{minipage}[t]{0.49\textwidth}
\includegraphics[height=0.75\textwidth]{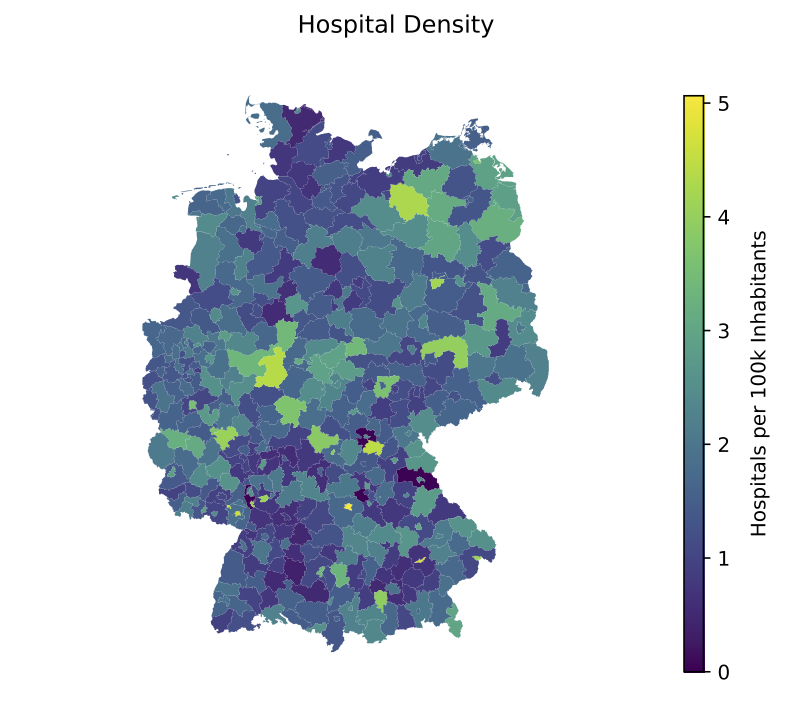}
\caption{\label{fig:hospitals} Overview the hospital density used within the \JuneGermany framework}
\end{minipage}
\end{center}
\end{figure*}

An overview of the distribution of households, e.g. what percentage are single, couple, with or without children, for three selected municipalities is given in Figure \ref{fig:households}. Again, there are significant differences between larger and smaller municipalities, which are important for modelling the evolution of infectious diseases. 

\subsection{Schools, Universities, Hospitals, and Care-Homes}

There is no central database for all schools in Germany, as school data is collected independently for each state. For some states, this includes not only the location of the schools but also the exact number of teachers and students. For those states where this data is not publicly available, the number of teachers and students is imputed using the average of the other states. Universities are modelled using information from various Wikipedia pages and public university and government websites. A total of 14,502 primary schools are modelled, with an average of 204 students per school. Secondary schools, of which 13,068 are included in the \JuneGermany framework, have a significantly higher average number of students per school, i.e. 506. An average teacher-to-student ratio of 0.12 and class sizes between 20 and 30 students are assumed.\par
A total of 418 universities and polytechnics are simulated, with associated student numbers ranging from a few hundred to several tens of thousands, giving an average of 6600 students per university. The size of university classes is assumed to be 200 students, taking into account that not all students attend the same classes and that there is a greater mix of students than in primary and secondary schools.

OpenStreetMaps is used for the location of hospitals, while information on the capacity of regular and ICU beds is gathered from \cite{coronadatenplattform}.

Only the number of standard and ICU beds per super area is available, not for each individual hospital. In order to distribute the number of standard and ICU beds, hospitals were first grouped by super area. The hospital beds were then distributed equally among them. A similar approach was used for ICU beds, but these were only distributed to hospitals with ICU capacity. 

\subsection{Work Places}

Jobs are categorized into different sectors according to the International Standard Industrial Classification. To create the companies in each district per sector, the average number of employees for a company in the specific sector is multiplied by the number of people working in that sector in the district. While this will not be perfectly accurate for all sectors, we assume that it is sufficiently accurate. When generating the population, each person is assigned a job either in the super area of residence or in a neighboring super area, based on the mobility data discussed in \autoref{sec:mobility}. The distribution of workers across the five sectors out of a total of 21 is shown in \autoref{fig:work-sectors}, also divided by gender. This categorization is highly important for the modelling of mitigation policies against the spread of the virus, as it allows for studying the impact of the closure of selected workplaces.

\begin{figure*}[htb]
\begin{center}
\begin{minipage}[t]{0.49\textwidth}
\includegraphics[height=0.75\textwidth]{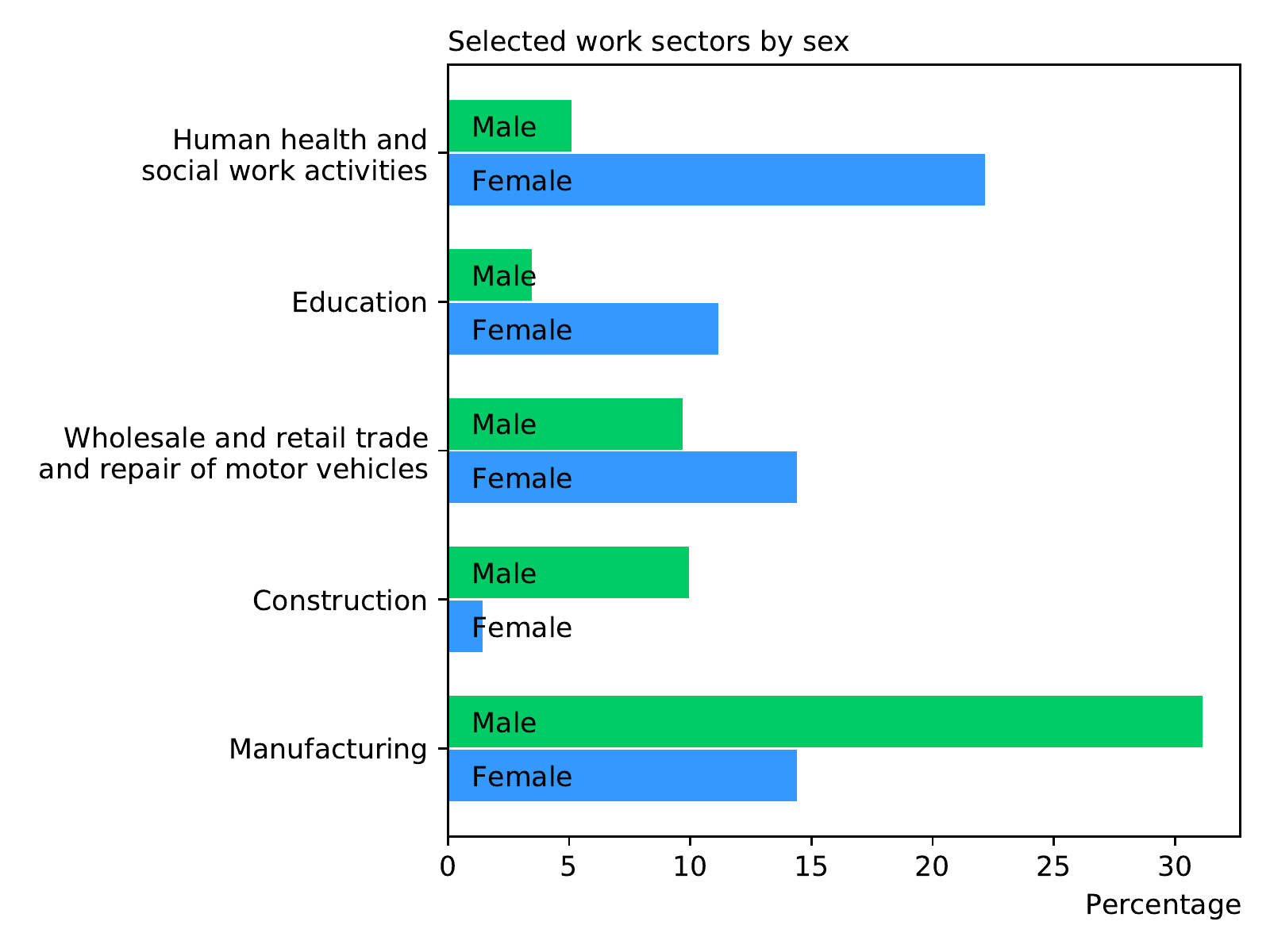}
\caption{\label{fig:work-sectors} Overview selected workplaces or work categories, separated by male and female.}
\end{minipage}
\hspace{0.1cm}
\begin{minipage}[t]{0.49\textwidth}
\includegraphics[height=0.75\textwidth]{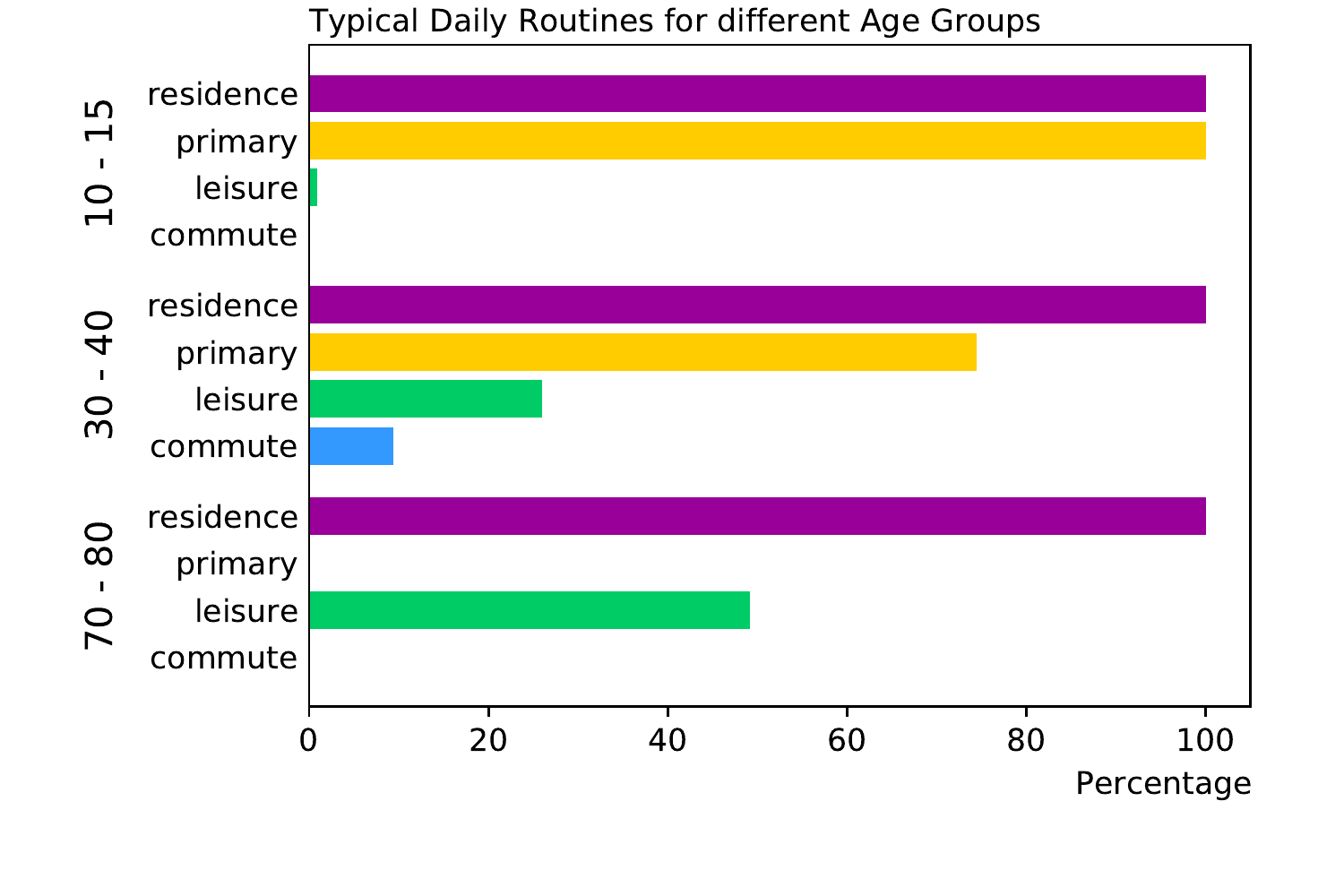}
\caption{\label{fig:daily} Overview of the daily activities in percent of 24h for selected age groups.}
\end{minipage}
\end{center}
\end{figure*}

\subsection{Social Activities and Interactions}

The modelling of social activities and interactions is a crucial part of the \June framework, as most transmissions take place in this context. These features of \June v1.0 have not been changed.\par
The agents' daily routines on weekdays are divided into four distinct activities: work/school, shopping, leisure, and staying at home. Leisure activities outside of working hours range from going to the cinema or theatre to meeting friends in pubs and restaurants. Shopping activities can be either grocery stores or retail shops. The location and distribution of the different leisure venues is determined using publicly available data from OpenStreetMaps.\par

Each venue is associated with a typical meeting duration of agents and typical group sizes, e.g. they are significantly larger for cinema visits compared to shopping in grocery stores. The time spent on the different activities depends mainly on the age of the agent. An overview of the main activities for three different age groups within the simulation is shown in \autoref{fig:daily}, with no mitigation policies applied. This is an important internal validation test, as the activities are taken directly from the simulation and not from the underlying data. For example, agents in the 70-80 age group spend no time at work, while increasing their time at home.

\subsection{Mobility\label{sec:mobility}}
The granularity of \June also lends itself to modelling travel patterns \cite{Bullock:2021}. Commuting is implemented by a directed network graph, where the edges correspond to transit routes and the nodes are major transit cities in Germany. In total, we select 15 major transit cities, corresponding to the most frequented train stations~\cite{wikiTravel}. Travellers move between these nodes and may share their means of transport.\par
\June distinguishes between two types of transport, public and private. Infection via private transport is assumed to be negligible, so only public transport is modelled, and June assigns each agent to either public or private transport. The spatial distribution of transport modes is taken from data provided by the German Federal Ministry of Transport and Digital Infrastructure~\cite{mid2017}.\par
\June also distinguishes between external and internal commuters. External commuters are those who live in non-metropolitan regions and commute to a metropolitan region. Internal commuters live and work in the same metropolitan region. Data on commuting in Germany are taken from the Federal Employment Agency~\cite{pendler}. 

\begin{figure*}[htb]
\begin{center}
\begin{minipage}[t]{0.49\textwidth}
\includegraphics[height=0.75\textwidth]{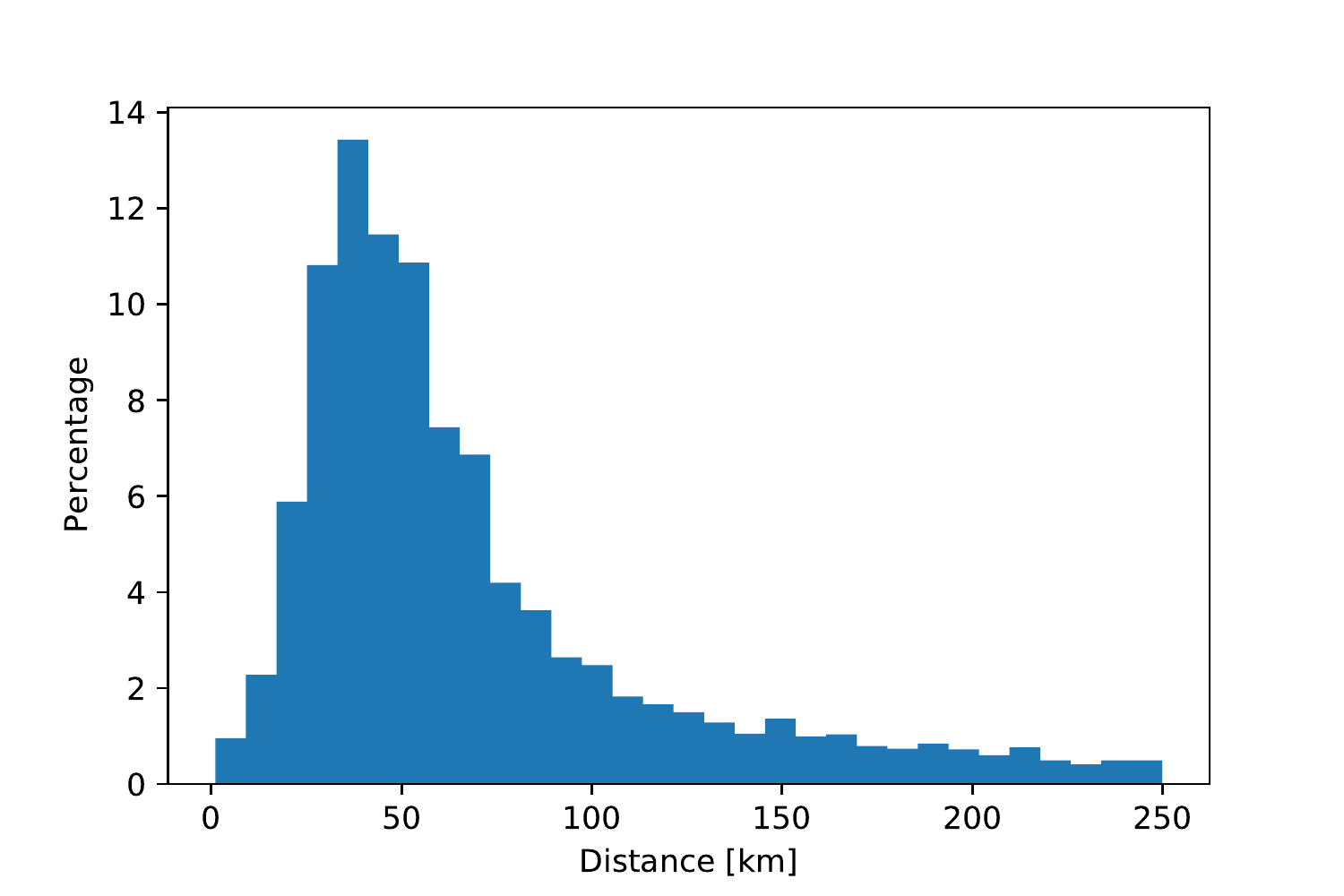}
\caption{\label{fig:work-sectors} Distribution of commuting distances of agents within Germany (Cutoff by 250km) before any restriction measures have been implemented.}
\end{minipage}
\hspace{0.1cm}
\begin{minipage}[t]{0.49\textwidth}
\begin{center}
\includegraphics[height=0.70\textwidth]{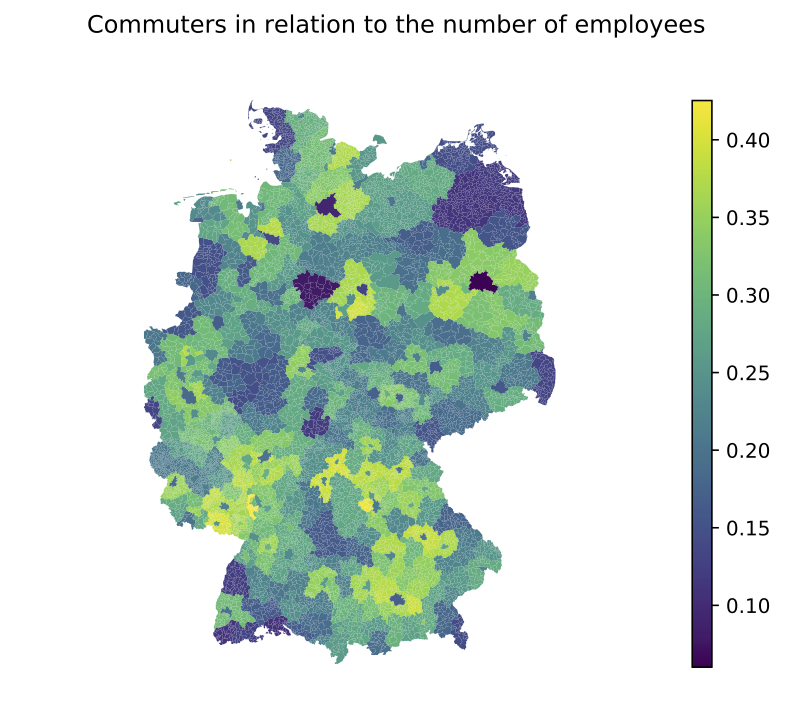}
\end{center}
\caption{\label{fig:outcommuter} Out-commuters in relation to the number of employees per super area. We consider the 20 super areas in which most commuters work (max. 250km away).}
\end{minipage}
\end{center}
\end{figure*}

The fraction of people commuting out of each super region is shown in Fig.~\ref{fig:outcommuter}. For the metropolitan regions, it can be seen that the number of people commuting out of the super areas is small. Therefore, in order to reduce the computation time, we only model commutes that start outside or within the metropolitan areas.\par
Internal commuters are modelled independently of their actual movement within the city. \June assigns them to groups with which they can interact. For external commuters, \June identifies common routes for commuters living in neighboring areas. The number of possible routes for each metropolitan area is set to four. Commuters are allocated to the available routes and commuters sharing a route have a probability of interacting.\par
Currently, only adult commutes to and from work are considered, due to a lack of information on student commuting patterns.

The mode of transport used by agents within \JuneGermany distinguishes between public transport, private transport, or no transport at all in the case of home-office arrangements, while it differs depending on the actual location. While in large cities, e.g. Berlin, a rather high proportion of agents use public transport (25\%), this proportion decreases to a small number in more rural areas.

\subsection{Policies\label{sec:policies}}

The \textit{policies} for \June v1.0 have been implemented to include both government policy and compliance. For example, \textit{mask wearing} is implemented to be either on or off depending on the date and has a compliance percentage associated with it. In the modification of these policies to \JuneGermany we do not alter this ability to include behavioral data. We will also deactivate any policies that do not apply to Germany. The main sources used to track the dates of both federal and state policies within Germany, as well as measures taken by the Robert Koch Institute (RKI), were Refs.~\cite{chronik,Archiv}. These sources provided the information needed to implement the \textit{social distancing}, \textit{mask wearing}, \textit{closure of schools}, \textit{closure of universities}, \textit{closure of leisure venues} and \textit{quarantine} policies. For \JuneGermany we consider \textit{hospitalization} and \textit{stay at home} policies to be constant, as was the case for \June v1.0. \par
The parameters \textit{regional compliance} and \textit{tiered lockdown} were not considered for \JuneGermany due to the variable and limited data available for individual German states. Instead, these parameters were set globally for Germany. The period in which the policies and compliance were non-uniform across German states was both limited and almost exclusively outside of all 3 major infection waves which occurred prior to publication. Therefore, this decision is expected to have a limited effect on the accuracy of our results. \textit{Limiting long commutes} and \textit{shielding} were not federal or state government policies in Germany and are therefore deactivated in this model. For more information on age-related behavior during the pandemic, see Ref.~\cite{lages2021relation}. \par
All other variables implemented in the policies are based on behavioral data from a range of sources that provided the most accurate publicly available data for these probabilistic parameters at the time of development. It is expected that the accuracy of these parameters could subsequently be improved as a wider range of data sets become available. In Germany, the closure of physical operations was not mandated or statistically monitored in the same way as in the UK. It was, therefore, necessary to use behavioral data on home office use from \texttt{statista} to apply the \textit{company closures} variable \cite{statista_homeoffice}. The \textit{leisure change probability} is probabilistic, so it was straightforward to implement using data from the \texttt{Gutenberg Covid-19 Study}. \cite{GCS}


\section{Estimation of Model-Parameters\label{sec:parameters}}

Due to the complexity of the JUNE simulator, particularly the large dimension of both input parameter and output space, emulation and history matching \cite{hme1, hme2} was used to try to find acceptable matches to data. An \emph{emulator} is a statistical surrogate for a model: suppose we have a single output $f(x)$ from the simulator given input parameters $x\in\mathbb{R}^d$. Then the equivalent emulator is
\begin{equation}
g(x) = \sum_{i=1}^p \beta_i h_i(x) + u(x),
\end{equation}
where $h_i(x)$ are a collection of simple functions of $x$, $\beta_i$ the corresponding coefficients, and $u(x)$ a weakly stationary second-order process which encodes the influence a point $x$ has on neighboring points in input space. Full distributional specifications can be chosen for the (unknown) coefficients $\beta_i$ and the form of $u(x)$; we instead take a pragmatic approach and follow the Bayes linear paradigm \cite{bayeslinear} which requires specification of only expectations, variances, and covariances. This approach is helpful when we do not have the means to specify distributions for all quantities or, even if full specification is possible, we do not necessarily believe in the validity of such a specification. Given the relevant choices for $\mathbb{E}[\beta]$, $\text{Var}[\beta]$, $\mathbb{E}[u(x)]$, $\text{Cov}[u(x), u(x^\prime)]$, and $\text{Cov}[\beta, u(x)]$ for any parameter sets $x,\, x^\prime$, we may update $g(x)$ with respect to runs $D=(f(x_1), \dots, f(x_n))$ from the simulator to produce a prediction of the simulator's output at any unseen point $x$, which we denote $\mathbb{E}_D[g(x)]$. Such predictions can be calculated in fractions of a second, offering orders-of-magnitude efficiency improvements on directly evaluating model output with recourse to the simulator.

The emulator is a statistical approximation to the output of the model, but any predictions come with an associated uncertainty $\text{Var}_D[g(x)]$ which is influenced by the prior specifications for the emulator and the proximity of the unseen point $x$ to any `training' points $x_i$. This is leveraged by the \emph{history matching} framework, which seeks to find acceptable parameter matches by complementarity. We define an \emph{implausibility measure}, $I(x)$, for a parameter set $x$ given observation $z$
\begin{equation}
I^2(x) = \frac{(\mathbb{E}_D[g(x)] - z)^2}{\text{Var}_D[g(x)] + \sigma_o^2 + \sigma_m^2},
\end{equation}
where here $\sigma_o^2$ and $\sigma_m^2$ represent any other uncertainties about the accuracy of our observational data and simulator output, respectively. The implausibility can be small either if the emulator prediction $\mathbb{E}_D[g(x)]$ is close to the observed value $z$ (representing a `good' fit to data), or if the uncertainties are large (representing a part of parameter space that would merit further investigation). For multiple outputs, we may define composite implausibility measures that provide a single measure of suitability across all outputs in question.

The history matching approach then seeks to rule out parts of parameter space that have low implausibility; that is points that, even accounting for the uncertainties in emulation, model performance, and observational error, cannot be close to our observation. Points thus generated can be entered into the simulator and used to construct more accurate emulators over the reduced parameter space, which in turn define a more accurate implausibility measure. This process can be continued until a large enough set of acceptable matches to data has been found; the emulator uncertainty is much smaller than any other sources of uncertainty; or available computational resources have been expended.

The emulation and history matching framework offers a number of advantages over traditional methods of parameter estimation. It requires relatively few evaluations from the (computationally expensive) simulator to train an emulator; emulator predictions are far more computationally efficient than direct evaluation for the simulator and therefore allow a more complete exploration of the full parameter space; the method offers a means of codifying and classifying sources of discrepancy between our simulated output, observations, and real-world phenomena; and the complementarity of the approach means that the final space identified comprises all possible acceptable points for matching to our data, rather than a single point that may not be representative of the `true' parameter values in reality. To utilise this approach, we used the in-development \texttt{R} package \texttt{hmer} \cite{hmer}, which streamlines the process of constructing emulators and generating representative parameter sets for later iterations of emulation and history matching, and has been used for parameter estimation in other epidemiological scenarios.


\section{Validation and Results\label{sec:validation}}

The validation of \JuneGermany is carried out using data from the state of Rhineland-Palatinate between October 2020 and February 2021, i.e. during the second wave. The number of cumulative deaths in the age groups 0-4, 5-14, 15-34, 35-59 and $\geq$60 from the 1st of October 2020 to the 14th December 2021 were used to determine the best model parameters\footnote{In fact, simulations have been performed until the 25th of December, however, the best parameter choice was selected for the 14th December 2021.} (Section \ref{sec:parameters}). These model parameters are then used to simulate the full second wave until the 22nd of February 2022. The number of cumulative deaths was chosen as the fitting observable, as it appears to be the most reliable dataset available. 

\begin{figure*}[thb]
\centering
    \includegraphics[width=0.49\textwidth]{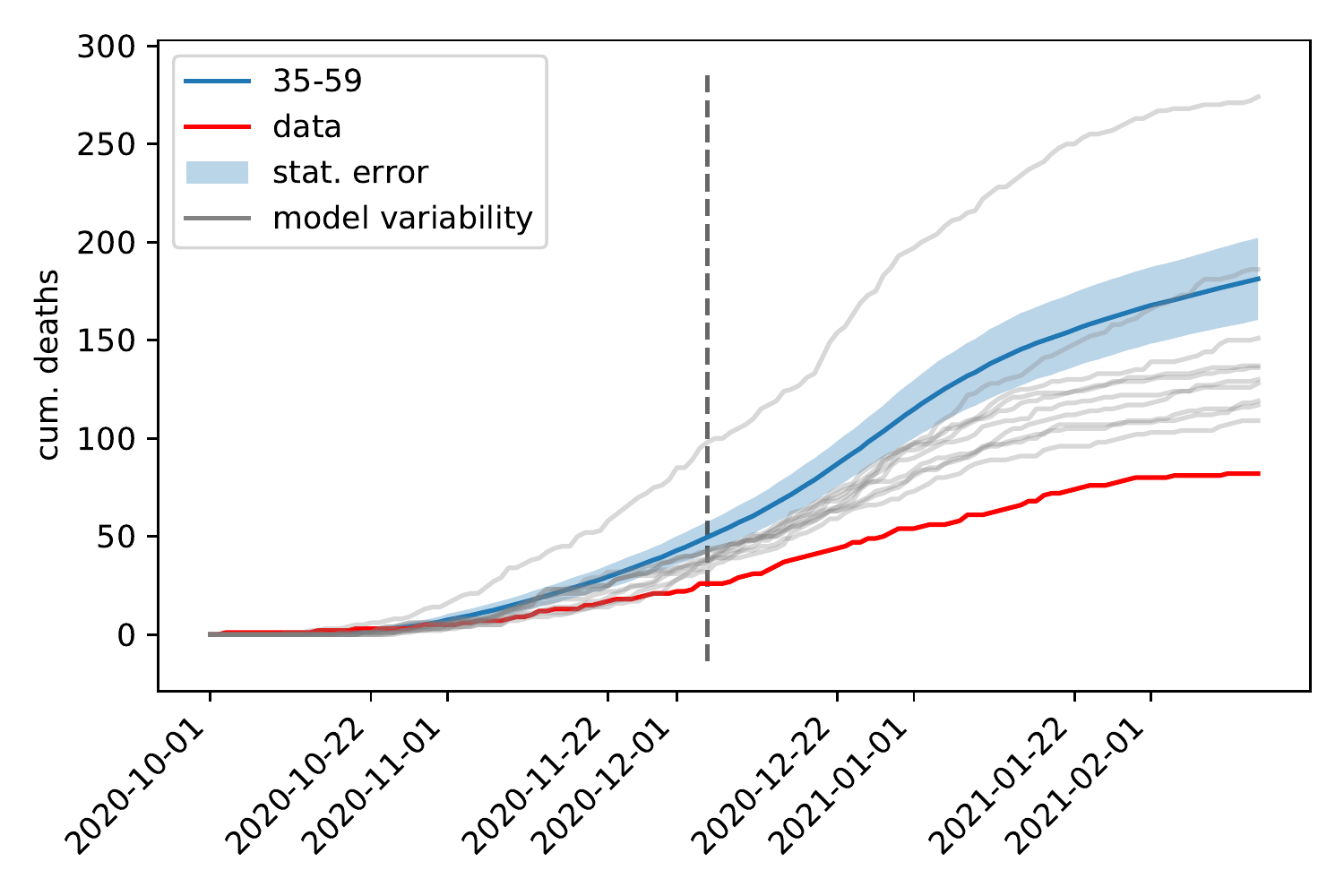}
    \includegraphics[width=0.49\textwidth]{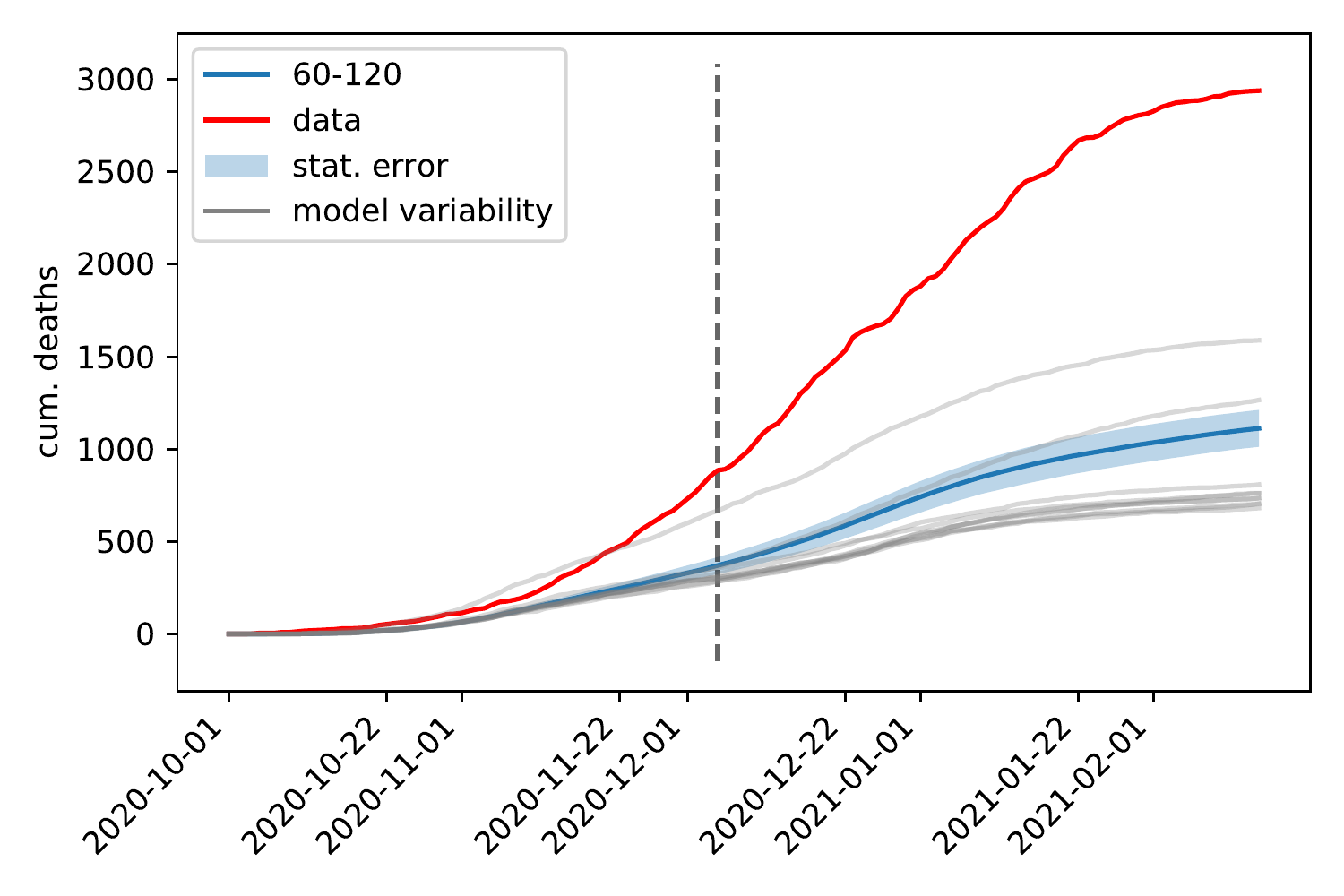}
    \caption{Recorded cumulated number of death (red line) for the age group 35-59 (left) and $\geq$60 (right), together with the \JuneGermany simulation with the best fitting parameters (blue line), its statistical uncertainty (shaded blue) and the simulated curves of alternative sets of model parameters tested during the fitting procedure (gray). The vertical line indicates the date until when the data was fitted.}
\label{fig:res1}
\end{figure*}

While most deaths are observed in the $\geq$60 age group, followed by the 35-59 age group, almost no deaths have been observed in the lower age groups. The fit is therefore mainly based on the distributions of the top two groups, shown in figure \ref{fig:res1}. The best fit is indicated by the blue line, while the blue band indicates the statistical uncertainty. The grey lines show simulated runs of different model parameters corresponding to 68\% of all simulated runs that come closest to the best fit. The chosen set of parameters gives too small values for the age group $\geq$60, while overestimating the number of deaths in the age group 35-59. This discrepancy could be reduced by using more iterations in the fitting procedure, but the correlation of the cumulative number of deaths between the two age groups makes it unlikely that a better fit will be observed. Another possible explanation is the poor quality of the available data in Germany, as the \June simulation framework provided a good description in England. 

\begin{figure*}[thb]
\centering
    \includegraphics[width=0.49\textwidth]{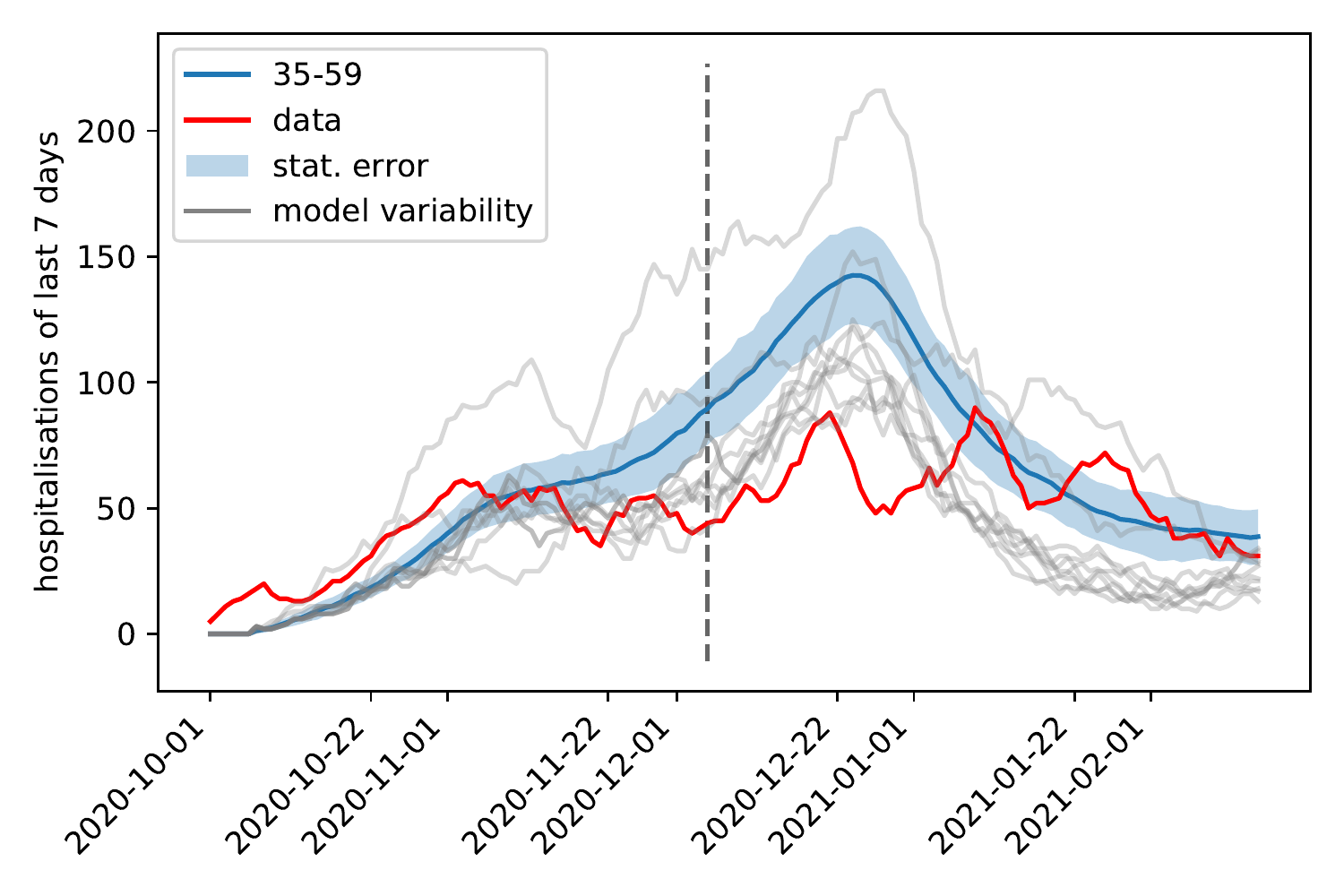}
    \includegraphics[width=0.49\textwidth]{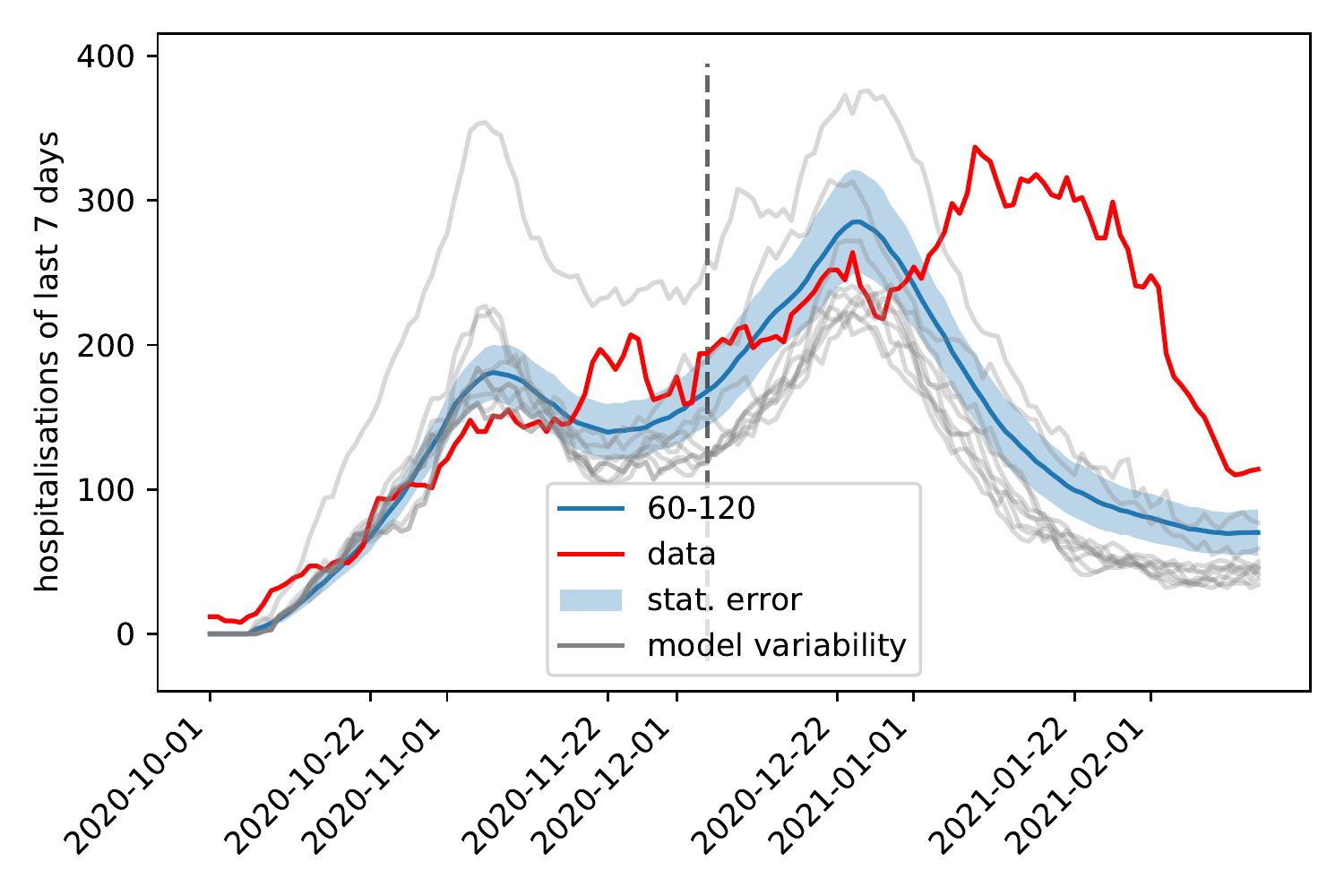}
    \caption{Hospitalization rate of the last 7 days  (red line) for the age group 35-59 (left) and $\geq$60 (right), together with the \JuneGermany simulation with the best fitting parameters (blue line), its statistical uncertainty (shaded blue) and the simulated curves of alternative sets of model parameters tested during the fitting procedure (gray). The vertical line indicates the date until when the data was fitted.}
\label{fig:res2}
\end{figure*}

Despite the shortcomings in predicting the actual number of victims at the end of the second wave (19 March 2021), which is 1550 across all age groups compared to 3195 cases in the official data, the simulation correctly predicts the length of the wave as well as its peak. The comparison of the hospitalization rate of the last 7 days is shown in Figure \ref{fig:res2}, where a good agreement can be observed, although the simulation predicts a faster decline of the wave than observed in the data. The total number of patients predicted to be hospitalized during the entire second wave is 5181, compared with the official number of 5638. 

\begin{figure*}[thb]
\centering
    \includegraphics[width=0.49\textwidth]{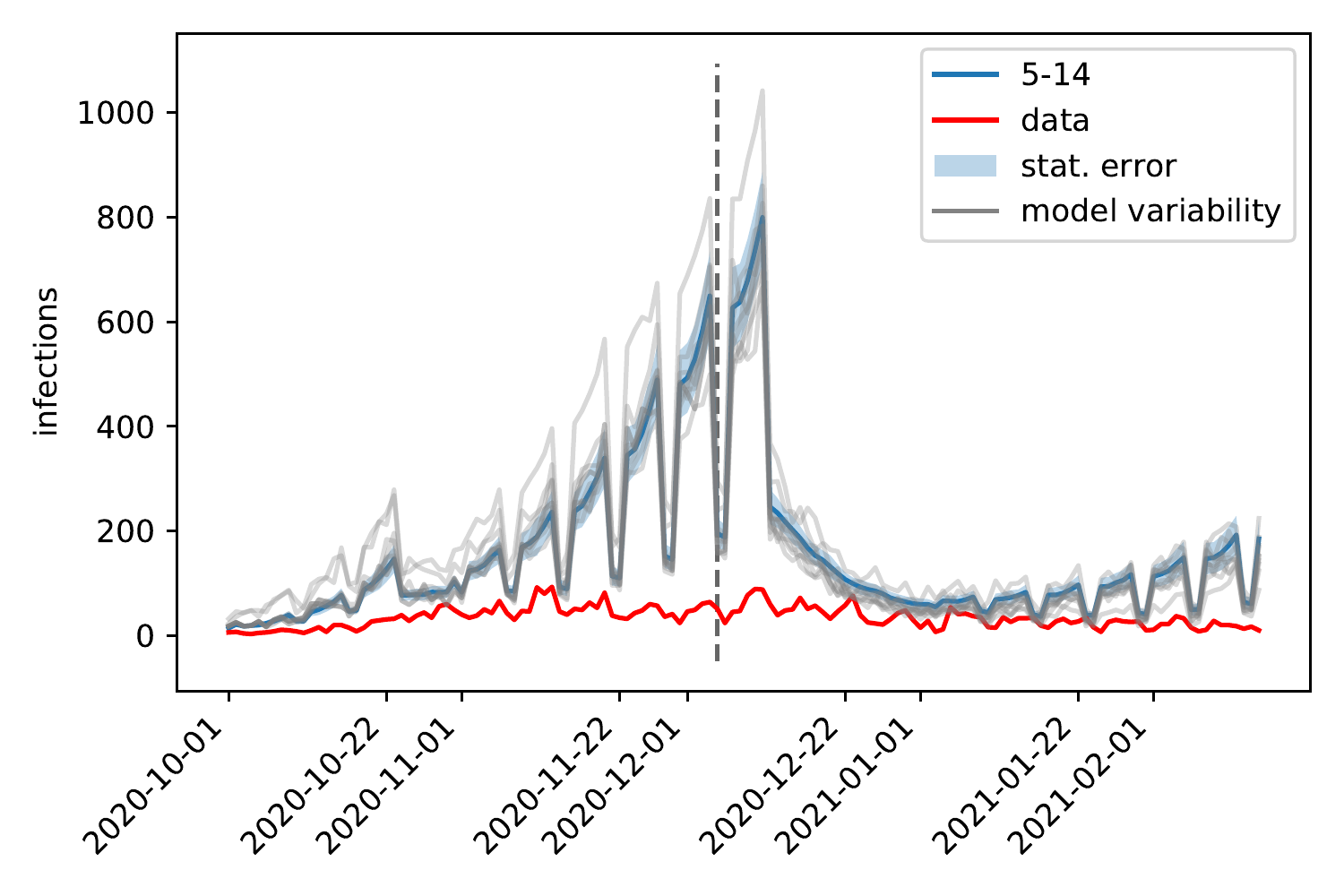}
    \includegraphics[width=0.49\textwidth]{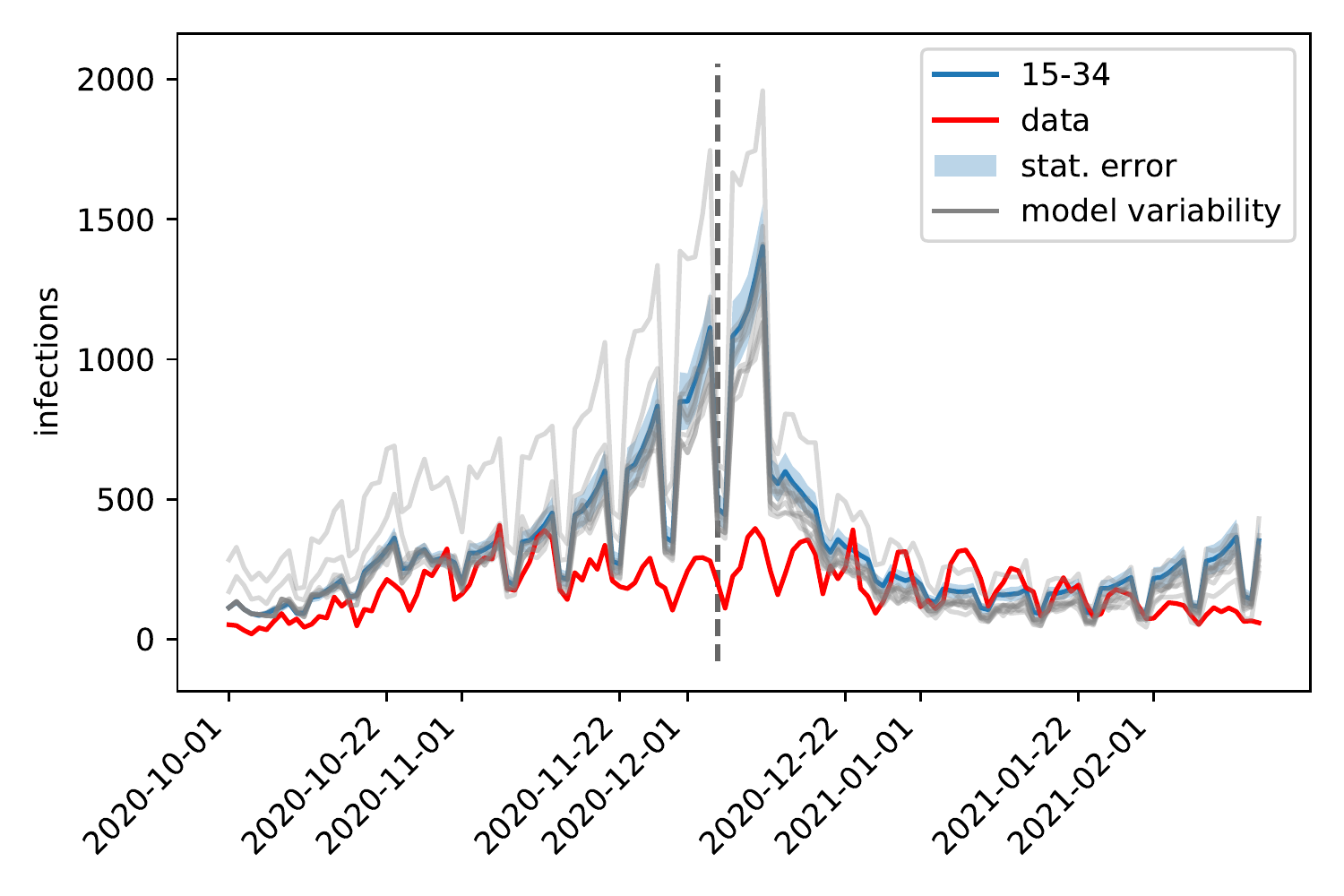}
    \includegraphics[width=0.49\textwidth]{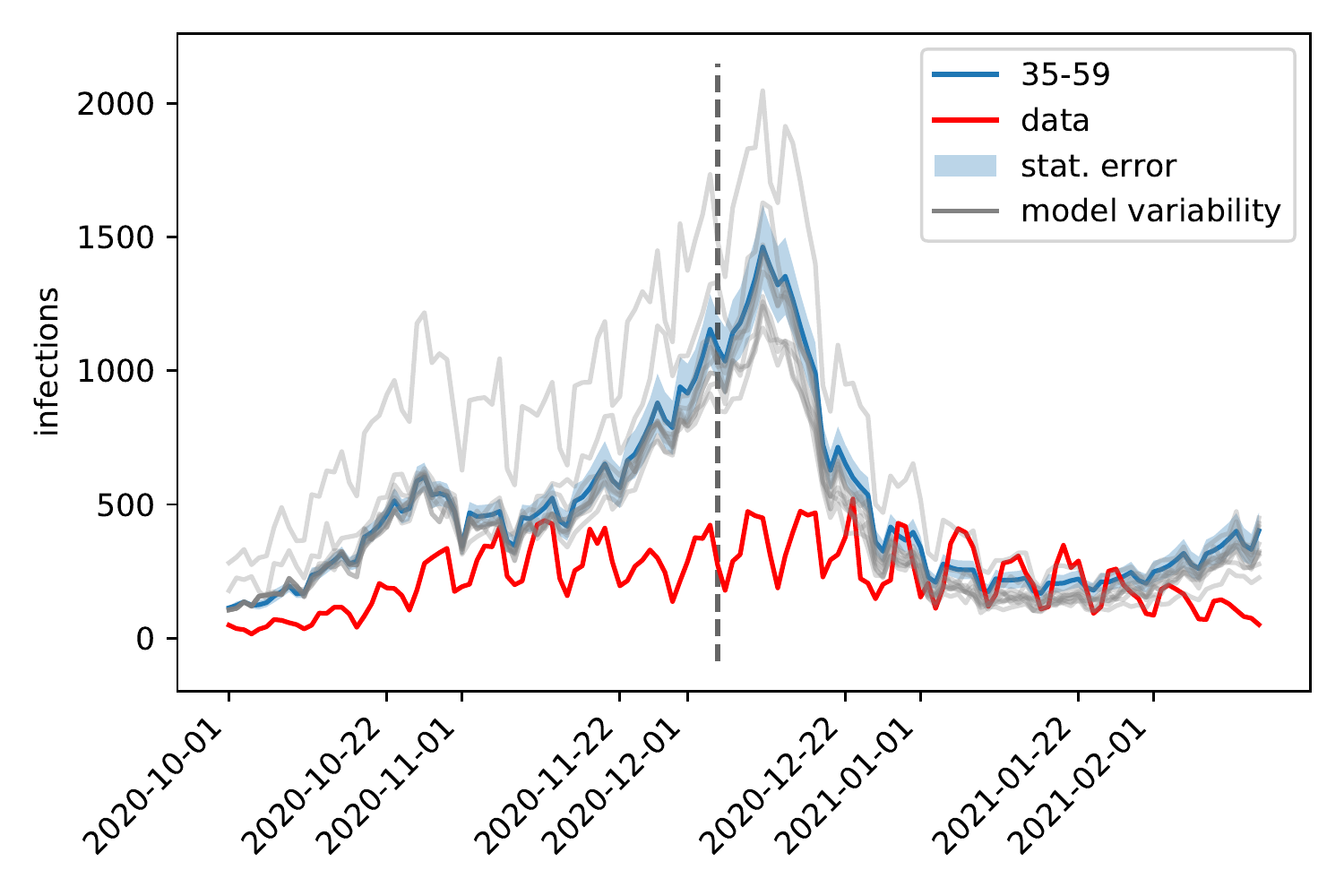}
    \includegraphics[width=0.49\textwidth]{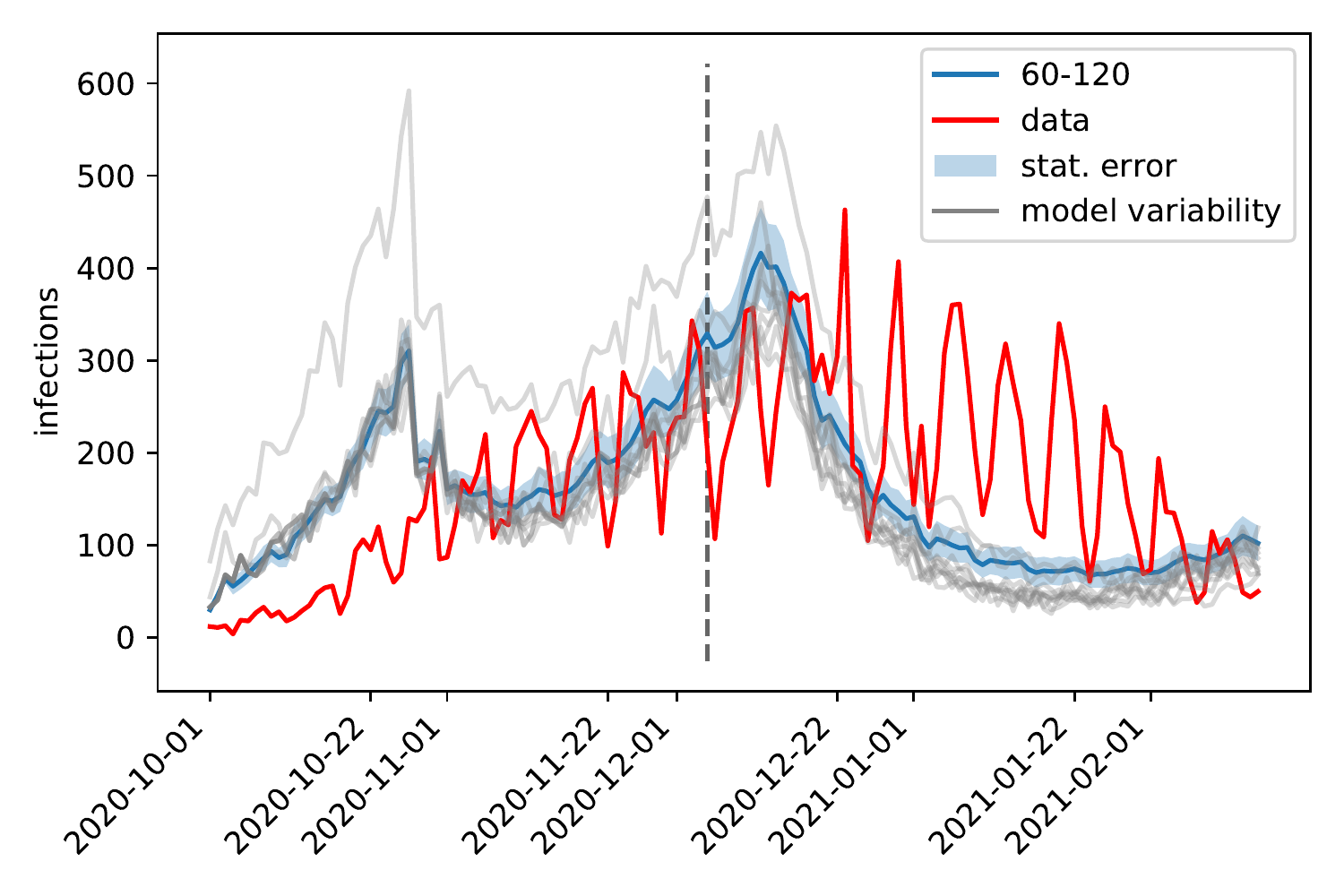}
    \caption{Number of infections (red line) for the age group 5-14 (upper left), 15-34 (upper right), 35-59 (lower left) and $\geq$60 (lower right), together with the \JuneGermany simulation with the best fitting parameters (blue line), its statistical uncertainty (shaded blue) and the simulated curves of alternative sets of model parameters tested during the fitting procedure (gray). The vertical line indicates the date until when the data was fitted.}
\label{fig:res3}
\end{figure*}

The official incidence rates as well as the predictions of \JuneGermany for all age groups is shown in Figure \ref{fig:res3}. While the number of infections is correctly described for the age group $\geq$60, for the age groups 14-34 and 35-59 there is a discrepancy by a factor of about three, i.e. a number of unreported cases. This ratio increases significantly for the 5-14 age group. In fact, this trend is to be expected, as older people are more likely to suffer from severe symptoms and large testing campaigns have been carried out in rest homes, while Covid-19 infection in younger people is more likely to go undetected.

\section{Conclusion\label{sec:conclusion}}

\JuneGermany is a new agent-based simulation of the spread of diseases within the population of Germany, which allows the modelling of testing and vaccination strategies. It is possible to model several virus variants simultaneously within \JuneGermany. The published code contains a complete geographical, demographic, and sociological description of the German population at an unprecedented level of detail and is capable of simulating the behaviour of more than 80 million agents. The model and its data have been validated for the case of the state of Rhineland-Palatinate, where reasonable agreement between simulation and reported data is observed. 

\section*{Acknowledgement}

The authors would like to thank Frank Krauss and Joseph Aylett-Bullock of Durham University for their help during our further developments for \JuneGermany. The authors would also like to thank Artur Barcsay, Michael Käfer, and Philip Kennedy from the Johannes Gutenberg University of Mainz for their work on earlier versions of the \JuneGermany framework. This work was supported by the Johannes Gutenberg Startup Research Fund and would not have been possible with the ERC grant LightAtLHC. Part of the simulations were performed on the Mogon II supercomputer at the Johannes Gutenberg University Mainz (hpc.uni-mainz.de). The authors gratefully acknowledge the computing time granted on the supercomputer.

\bibliographystyle{elsarticle-num}
\bibliography{Bibliography}

\end{document}

%% file: author_list.tex
\author{Kerem Akdogan\inst{1*}, Lucas Heger\inst{1*},  Andrew Iskauskas\inst{2}, Friedemann Neuhaus\inst{1} \and Matthias Schott \inst{1,3}}
\institute{%
    \inst{1} Institute of Physics, Johannes Gutenberg University, Mainz, Germany\\
    \inst{2} Department of Mathematical Sciences, Durham University, United Kingdom\\
    \inst{3} Institute of Computer Science, Philipps-University, Marburg, Germany\\
    \inst{*} Corresponding authors \\
}